%% file: dihedral_benchmarking.tex
\documentclass[aps,english,10pt,nofootinbib,superscriptaddress,notitlepage,twocolumn]{revtex4-1}

\include{Qcircuit}
\include{macros}

\usepackage{graphicx}

\begin{document}
	\title{Characterizing Universal Gate Sets via Dihedral Benchmarking}
	\author{Arnaud Carignan-Dugas}
	\affiliation{Institute for Quantum Computing and the Department of Applied
		Mathematics, University of Waterloo, Waterloo, Ontario N2L 3G1, Canada}
	\author{Joel Wallman}
	\affiliation{Institute for Quantum Computing and the Department of Applied
		Mathematics, University of Waterloo, Waterloo, Ontario N2L 3G1, Canada}
	\author{Joseph Emerson}
	\affiliation{Institute for Quantum Computing and the Department of Applied
		Mathematics, University of Waterloo, Waterloo, Ontario N2L 3G1, Canada}
	\affiliation{Canadian Institute for Advanced Research, Toronto, Ontario M5G 1Z8, Canada}
	\date{\today}
	\begin{abstract}
		We describe a practical experimental protocol for robustly 
		characterizing the error rates of non-Clifford gates associated with 
		dihedral groups, including gates in SU(2) 
		associated with arbitrarily small angle rotations. Our dihedral 
		benchmarking protocol is a generalization of randomized benchmarking 
		that relaxes the usual unitary 2-design condition. 
		Combining this protocol with existing randomized benchmarking schemes enables an efficient means of characterizing universal gate sets for 
		quantum information processing in a way that is independent of 
		state-preparation and measurement errors.  In particular, our protocol 
		enables direct benchmarking of the $T$ gate (sometime called 
		$\pi/8$-gate) even for the gate-dependent error model that is expected 
		in leading approaches to fault-tolerant quantum computation. 
	\end{abstract}
	
	\maketitle
	
	A universal quantum computer is a device allowing for the implementation of 
	arbitrary unitary transformations. As with any scenario involving control, 
	a practical quantum computation will inevitably have errors. While the 
	complexity of quantum dynamics is what enables the unique 
	capabilities of quantum computation, including important applications such 
	as quantum simulation and Shor's factoring algorithm, that same complexity 
	poses a unique challenge to efficiently characterizing the errors. One approach is quantum 
	process tomography~\cite{Chuang1997,Poyatos1997}, which yields an 
	informationally-complete characterization of the errors on arbitrary 
	quantum gates, but requires resources that scale exponentially in the 
	number of qubits. Moreover, quantum process tomography can not 
	distinguish errors associated with the quantum gates from those associated with 
	state-preparation and measurement (SPAM)~\cite{Merkel2013}. 
	
	Randomized 
	benchmarking~\cite{Emerson2005, Knill2008, Magesan2011, Magesan2012a} using 
	a unitary 2-design~\cite{Dankert2006,Dankert2009}, such as the Clifford group, 
	overcomes both of these limitations by providing an estimate of the error 
	rate per gate averaged over the 2-design. More specifically, it is a method 
	for efficiently estimating the average fidelity 
	\begin{equation}
	\mc{F}_{\rm{avg.}}(\mc{E}) := \int d\psi \langle \psi |\mc{E}(\psi) |\psi
	\rangle
	\end{equation}
	of a noise map $\mc{E}$ associated with any group of quantum operations 
	forming a unitary 2-design in a way that is independent of SPAM errors. This partial information is useful in practice 
	as it provides an efficient means of tuning-up experimental performance, 
	and, moreover, provides a bound on the threshold error rate required for 
	fault tolerant quantum computing \cite{Gottesman2010}, a bound that becomes tight 
	when the noise is stochastic \cite{Magesan2012a,Sanders2015}. 
	
	An important limitation of existing randomized benchmarking methods is that they 
	are only efficient in the number of 
	qubits~\cite{Emerson2005,Dankert2009,Magesan2011} for non-universal sets of gates such as the Clifford group. While Clifford gates play an important role in leading approaches to fault-tolerant quantum 	computation based on stabilizer codes~\cite{Gottesman2010}, one still needs a means of benchmarking the remaining non-Cifford gates required for universality; this is particularly important because the non-Clifford gates will be implemented via magic state distillation and gate injection \cite{Bravyi2005, Meier2012}, which is a complex procedure  that will be subject to dramatically different error rates than those of the (physical or logical) Clifford gates.
	
	In the present paper, we describe a protocol for benchmarking the 
	average fidelity of a group of operations corresponding to the dihedral 
	group which does not satisfy the usual 2-design constraint for randomized 
	benchmarking. However, we show that the dihedral benchmarking protocol 
	still allows the average fidelity to be estimated while retaining the 
	benefits of standard randomized benchmarking. Furthermore, by combining 
	our dihedral benchmarking protocol with both standard~\cite{Magesan2011} 
	and interleaved randomized benchmarking~\cite{Magesan2012b}, we give an 
	explicit method for characterizing the average fidelity of the $T$ gate 
	directly. This is of particular interest because the $T$ gate combined with 
	the generators of the Clifford group (e.g., the CNOT, the Hadamard and the 
	Pauli gates) provides a standard gate set for generating universal quantum 
	computation. 
	Moreover our protocol enables characterization of non-Clifford gates associated with arbitrarily small angle rotations, which are of interest to achieve more efficient fault-tolerant circuits~\cite{ Landahl2013, Forest2015, Duclos-Cianci2015}.
Remarkably, our protocol overcomes the key assumption of `weak gate-dependence' of the noise that limits previous  benchmarking protocols. Specifically,  the protocol is robust in the important setting when the error on the non-Clifford gate, such as the $T$ gate, is substantially different from the error on the Clifford operations.  As noted above, this is the expected scenario in leading approaches to fault-tolerant quantum computation. 

	\begin{figure}
		\centering
		\includegraphics[width= .5\linewidth]{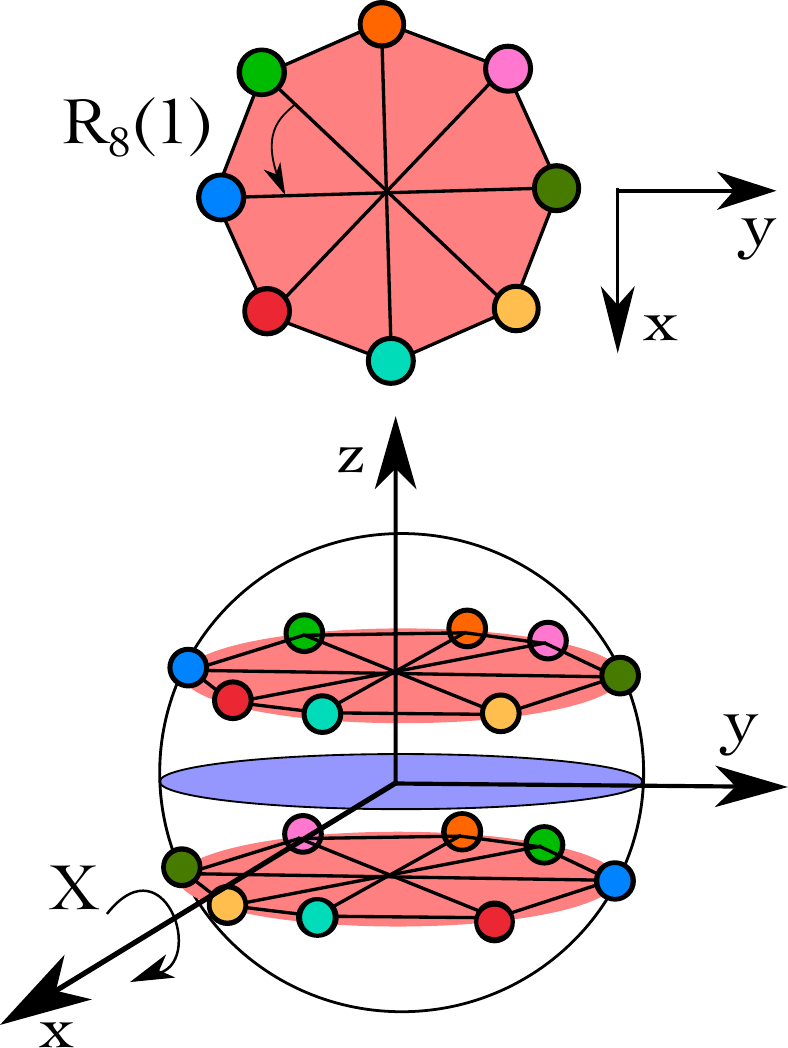}
		\caption{\label{fig:graph_class}
			The orbit, under the action 
			of the dihedral group $\mc{D}_8$, of an input state located at a $45^\circ$ 
			degree latitude on the Bloch sphere. $R_8(z)$ are the rotations of the octagon, while $X$ is a
			reflection (or a rotation in $3$ dimensions, with the rotation axis parallel
			to the octagon's surface). The $T$ gate corresponds to the smallest rotation
			$R_8(1)$.}
	\end{figure}
	
	\textit{Characterizing single-qubit unitary groups.}---We now outline a 
	protocol that yields the average gate fidelity of the experimental 
	implementation of a single-qubit unitary group of the form
	\begin{align}\label{eq:defdj}
		\mc{D}_j=\langle R_j(1),X \rangle,
	\end{align}
	where $j$ is a positive integer and
	\begin{align}\label{eq:rot}
		R_j(z):=e^{\pi i z Z/j}=\cos(\pi z/j)\unit +i \sin(\pi z/j) Z.
	\end{align}
	Up to an overall sign, $\mc{D}_j$ is a representation of the dihedral group of 
	order $2j$, with $XR_j(z) = R_j(z+j)X$, which is not a unitary 2-design and 
	includes gates producing arbitrarily small rotations as $j$ increases. Note 
	that the choice of rotation axis is arbitrary, and that any single-qubit 
	gate can be written as $R_j(1)$ relative to some axis. Consequently, our 
	protocol will allow any single-qubit gate to be benchmarked. The Bloch 
	sphere representation of $\mc{D}_8$ acting on a qubit state is illustrated in 
	fig.~\ref{fig:graph_class}. This group contains the so-called $T$ gate, 
	which corresponds to the $R_8(1)$ rotation. 
	
	The dihedral benchmarking protocol is as follows.
	\begin{enumerate}
		\item Choose two binary strings of length $m$, ${\bf z} = (z_1,\ldots 
		,z_{m}) \in \mbb Z_j^m$ and ${\bf x} = (x_1,\ldots ,x_{m}) \in \mbb Z_2^m$ 
		independently and uniformly at random.
		\item Prepare a system in an arbitrary initial state 
		$\rho$.\footnote{The constants $A$ and $B$ appearing in 
		eqns.~\ref{eq:decay_curvep0} 
			and \ref{eq:decay_curvep1} depend on state preparation, as shown in 
			eqns.~\ref{eq:A}--\ref{eq:B2}. These constants may 
			be maximized by choosing an appropriate state preparation (and the 
			corresponding measurement). In particular, optimal states for 
			eqn.~\ref{eq:decay_curvep0} and eqn.~\ref{eq:decay_curvep1} are 
			$|0\rangle\langle0|$ and $|+\rangle\langle+|$ 
			respectively.} 
		\item At each time step $t = 1,\ldots,m$, apply $R_{j}(z_t)X^{x_t}$.
		\item Apply the inversion gate, defined as
		\begin{center}$G_{\rm inv.}:=X^{b_1}Z^{b_2}\prod_{t=1}^m [R_{j}(z_t)X^{x_t}]\ct$, \end{center} 
		where $b_1, b_2 \in \mbb Z_2$ are fixed by considerations below.
		\item Perform a POVM $\{E, \unit - E \}\rightarrow \{+1,-1\}$ for some
		$E \approx \rho$, to estimate the probability $q(+1|m,x,z,b_1,b_2)$ of outome $+1$. 
		\item Repeat steps 1-5 $k$ times, where $k$ is fixed by the requirement to estimate the survival probability 
		\begin{center}$\textrm{Pr}(m,b_1,b_2):=\av{2j}^{-(m+1)}\sum_{x,z} q(+1|m, x,z,b_1,b_2)$ \end{center}
		to a desired precision (see \cite{Magesan2012a, Granade2014, Wallman2014} for details on the required sampling complexity).
	\end{enumerate}
	For $b_1=b_2=0$, the average survival probability is
	\begin{align}\label{eq:decay_curve00}
		\textrm{Pr}(m,0,0) = A p_0^m + B p_1^m+C~,
	\end{align}
	where $A$, $B$ and $C$ are constants absorbing SPAM factors. 
	Because a sum of two exponentials leads to a non-linear fitting problem, it 
	should generally be advantageous to fit instead to
		\begin{align}\label{eq:decay_curvep0}
			 & \textrm{Pr}(m, 0, 0) + \textrm{Pr}(m, 0, 1) \notag \\
			 - & \textrm{Pr}(m, 1, 0) - \textrm{Pr}(m, 1, 1)=  4 A p_0^m
		\end{align}
and 
		\begin{align}\label{eq:decay_curvep1}
			\textrm{Pr}(m, 0, 0) -  \textrm{Pr}(m, 0, 1)=  2 B p_1^m.
		\end{align}
	As we will show below, the average gate fidelity is related to the fit 
	parameters $p_0$ and $p_1$ by
	\begin{align}\label{eq:ave_fid}
		\mc{F}_{\rm{avg.}}(\mc{E})=\frac{1}{2}+\frac{1}{6}(p_0+2p_1).
	\end{align}
	
	\emph{Analysis.}---We now derive the formula for the decay curves expressed 
	in eqns.~\ref{eq:decay_curve00}--\ref{eq:decay_curvep1}, together with 
	the average fidelity eqn.~\ref{eq:ave_fid}.
	 We assume that the experimental noise is
	completely positive and trace-preserving and is also gate and time-independent (though perturbative 
	approaches to relax these assumptions can be considered~\cite{Magesan2012a,Wallman2014}), 
	so that we can represent the experimental implementation of 
	$\mc{R}_j(z)\circ \mc{X}^x$ as $\mc E \circ \mc{R}_j(z)\circ \mc{X}^x$. We 
	use calligraphic font to  denote abstract channels (where, for a unitary 
	$U$, the abstract channel $\mc{U}$ corresponds to conjugation by $U$) and 
	$\circ$ to denote channel composition (i.e., $\mc{B}\circ\mc{A}$ means 
	apply $\mc{A}$ then $\mc{B}$). We refer to
	\begin{align}
		\mc{E}^{\mc{G}} = (|\mc{G}|)^{-1}\sum_{\mc U\in\mc{G}} \mc 
		U^{-1}\circ\mc{E}\circ \mc U
	\end{align}
	as the \textit{twirl} of $\mc{E}$ over a group $\mc{G}$. Averaged over all
	sequences $x$ and $z$ of length $m$, our protocol yields the following 
	effective channel
	\begin{align}\label{eq:channel}
		{\mc C}=\mc{E} \circ \mc X^{b_1} \mc Z^{b_2} \circ \left( \mc E^{\mc D_j} \right)^{\circ m}~.
	\end{align}
	
	The Pauli-Liouville representation (see, e.g., Ref.~\cite{Wallman2014} for
	details) of an abstract channel $\mc E$ consists in a matrix of inner
	products between Pauli matrices $P_j$ and their images $\mc E(P_k)$
	\begin{align}
		\bmc E_{jk}=\text{Tr}(P_j \mc E(P_k))~. 
	\end{align}
	We denote this representation with the bold font $\bmc E$.
	The Pauli-Liouville representation of $\mc{D}_j$ is a
	direct
	sum of three inequivalent irreducible representations (irreps) of the dihedral
	group:
	\begin{enumerate}
		\item $R_j(z) X^x \rightarrow 1$ (trivial~representation) 
		\item $R_j(z) X^x \rightarrow  
		\mat{cc}{
			\cos(2\pi z/j) & (-1)^{x+1}\sin(2\pi z/j) \\
			\sin(2\pi z/j) & (-1)^x\cos(2\pi z/j) 
		} $ (faithful~representation) 
		\item $R_j(z) X^x \rightarrow (-1)^x$ (parity~representation).
	\end{enumerate}
	This is easily seen by looking at the action of $\mc D_j$ on the Bloch 
	sphere (see fig.~\ref{fig:graph_class}). The trivial representation emerges
	from the unitality and trace-preserving properties of unitary operations, 
	which map any Bloch shell of constant radius to itself, including the 
	center point. The parity representation encodes the fact that the $\pm Z$ 
	poles of the Bloch sphere are invariant under conjugation by $R_j(z)$ and 
	swapped under conjugation by $X$. The two-dimensional representation 
	encodes the action of $R_j(z)X^x$ on the $XY$-plane of the Bloch sphere.
	
	As a consequence of Schur's lemmas (see the supplementary information of 
	Ref.~\cite{Gambetta2012}), the twirled noise channel is a direct sum of 
	three identity matrices:
	\begin{align}\label{eq:twirl}
		\bmc E^{\mc D_j}= \mat{cccc}{
			1 & 0 & 0 & 0\\
			0 & p_1 & 0 & 0\\
			0 & 0&p_1 & 0\\
			0 & 0 & 0 & p_0}~,
	\end{align}
	where $p_0 :=\bmc E_{44}$ and $p_1:=\frac{\bmc E_{22}+\bmc E_{33}}{2}$. 
	With these definitions, the average fidelity 
	is~\cite{Nielsen2002,Kimmel2014}
	\begin{align}		
		\mc{F}_{\rm{avg.}}(\mc{E})&=\frac{1}{2}+\frac{1}{6}({\bmc{E}}_{22}+{\bmc{E}}_{33}+{\bmc{E}}_{44})
		\notag\\
		&= \frac{1}{2} + \frac{1}{6}(p_0 +2 p_1)
	\end{align}
	as in eqn.~\ref{eq:ave_fid}. Using eqn. \ref{eq:twirl}, the effective channel $\bmc C$ from eqn. \ref{eq:channel} can readily be expressed as
		\begin{align}\label{eq:pauliC}
			\bmc C= \bmc E \cdot \mat{cccc}{
				1 & 0 & 0 & 0\\
				0 & (-1)^{b_2}p_1^m & 0 & 0\\
				0 & 0&(-1)^{b_1+b_2}p_1^m & 0\\
				0 & 0 & 0 & (-1)^{b_1}p_0^m}~.
		\end{align}
    Therefore the survival probability is
 		\begin{align}\label{eq:survexplicit}
 			& \textrm{Pr}(m, b_1,b_2)= \text{Tr} \left( E ~\mc C(\rho) \right) \notag \\
 			& = (-1)^{b_1} A p_0^m+\left((-1)^{b_1+b_2} B_1 +(-1)^{b_2} B_2\right) p_1^m +C~,
 		\end{align}
	where
 		\begin{align}
 			A~:= & ~2^{-1} \cdot \text{Tr}\left( E \cdot \mc E (Z)\right) \cdot  \text{Tr}\left(\rho Z\right)~, \label{eq:A} \\
 			B_1:= & ~2^{-1} \cdot\text{Tr}\left( E \cdot \mc E (Y)\right) \cdot \text{Tr}\left(\rho Y\right)~, \label{eq:B1}\\
 			B_2:= &~2^{-1} \cdot\text{Tr}\left( E \cdot \mc E (X)\right) \cdot \text{Tr}\left(\rho X\right)~,\label{eq:B2}\\
 			C~:= &~2^{-1} \cdot\text{Tr}\left( E \cdot \mc E (\mbb I )\right) ~.
 		\end{align}
	Eqns.~\ref{eq:decay_curve00}--\ref{eq:decay_curvep1} then follow from 
	appropriate choices of $b_1,b_2$ and simple algebra.
	
	\emph{Characterizing the $T$ gate.}---The $T$ gate, or the $R_8(1)$ 
	operation (see eqn. \ref{eq:rot}), is important in many implementations 
	since it is used to supplement the Clifford gates to achieve universal 
	quantum computation. In leading approaches to 
	fault-tolerant error-correction, the $T$ gate is physically realized via 
	magic-state injection \cite{Bravyi2005}, in which magic states are acted 
	upon by Clifford transformation and post-selected stabilizer measurements. Because the physical (logical) Clifford gates are applied directly (transversally) whereas the $T$ gate is implemented through the above method, 
 the error on the $T$ gate may be substantially different and requires separate characterization.  
	While the quality of the injected gate can be assessed by measuring the 
	quality of the input and output magic states as well as benchmarking the required stabilizer operations, here we provide a direct method to estimate the 
	average gate fidelity of the $T$ gate.	

	The $T$ gate is contained 
	in $\mc D_8$ (see eqn. \ref{eq:defdj}). $\mc D_8$ can be divided in two cosets: 
	$\mc D_4$ and $T \cdot \mc D_4$. $\mc D_4$, a subgroup of $\mc D_8$, is 
	generated by $X$ and the phase gate $S$, which are both Clifford operations. 
	If the average fidelity over $\mc D_8$ and $\mc D_4$ are similar, this is an indication 
	that the $T$ gate has similar average fidelity as the Clifford group. However, typically 
	this will not hold for the reasons stated above, in which case we suggest the following protocol. 
First benchmark $\mc D_4$ as per the above protocol.  Then adapt interleaved randomized
	benchmarking~\cite{Gambetta2012} to the above protocol by replacing steps $3$ and $4$  (with $j=4$) with the two following steps:
	\begin{itemize}
		\item[3$'$.] At each time step $t = 1,\ldots,m$, apply \newline 
		$R_{8}(1) 
		\circ R_{4}(z_t)X^{x_t}$.
		\item[4$'$.] Apply the inversion gate, defined as
		\begin{center}$G_{\rm inv.}:=X^{b_1}Z^{b_2}\prod_{t=1}^m [R_8(1)R_{4}(z_t)X^{x_t}]\ct$. \end{center} 
	\end{itemize}
	We require the sequence length to be even, so that the inversion gate is in 
	$\mc D_4$. Fitting the two decay curves obtained from 
	eqns.~\ref{eq:decay_curvep0} and~\ref{eq:decay_curvep1} allows the
	average fidelity $\mc{F}_{\rm{avg.}}(\mc{E}_T \circ \mc E)$ of the 
	composite noise map to be estimated via eqn. \ref{eq:ave_fid}. The average 
	fidelity of the $T$ gate, $\mc{F}_{\rm{avg.}}(\mc E_T)$, is then estimated 
	from the approximation $\chi_{00}^{\mc{E}_T \circ \mc 
	E}=\chi_{00}^{\mc{E}}\chi_{00}^{\mc E_T}$
	which is valid to within the implicit
	bound derived in~\cite{Kimmel2014}:
	\begin{align}\label{eq:implicit_bound}
		|\chi_{00}^{\mc{E}_T \circ \mc E}-\chi_{00}^{\mc{E}}\chi_{00}^{\mc E_T}| \leq
		&
		2\sqrt{(1-\chi_{00}^{\mc{E}})\chi_{00}^{\mc{E}}(1-\chi_{00}^{\mc{E}_T})\chi_{00}^{\mc{E}_T}}
		\notag \\
		&+(1-\chi_{00}^{\mc{E}})(1-\chi_{00}^{\mc{E}_T})~,
	\end{align}
	where in the qubit case
	\begin{align}
		\chi_{00}^{\mc{E}}=\frac{3}{2}\mc{F}_{\rm{avg.}}(\mc{E})-\frac{1}{2}~.
	\end{align}
This bound is loose in general but tight when the Clifford gates in $\mc D_4$ have much higher fidelity than the $T$ gate (which is the regime of interest  when optimizing the overhead and fidelity of the distillation and injection routines)~\cite{Fowler2012}.	

	\emph{Numerical simulation.}---Although the previous analysis is derived
	for gate- and time-independent noise, the randomized benchmarking 
	protocol is both theoretically and practically robust to some level of 
	gate-dependent noise~\cite{Magesan2012a,Wallman2014}. We now illustrate 
	through numerical simulations that this robustness holds for the dihedral 
	benchmarking protocol, particularly in the regime where the noise is 
	strongly gate-dependent (as expected when the gates are implemented using 
	different methods, namely, direct unitaries and magic state injection). 
	\begin{figure}
		\centering
		\includegraphics[width= 1.00\linewidth]{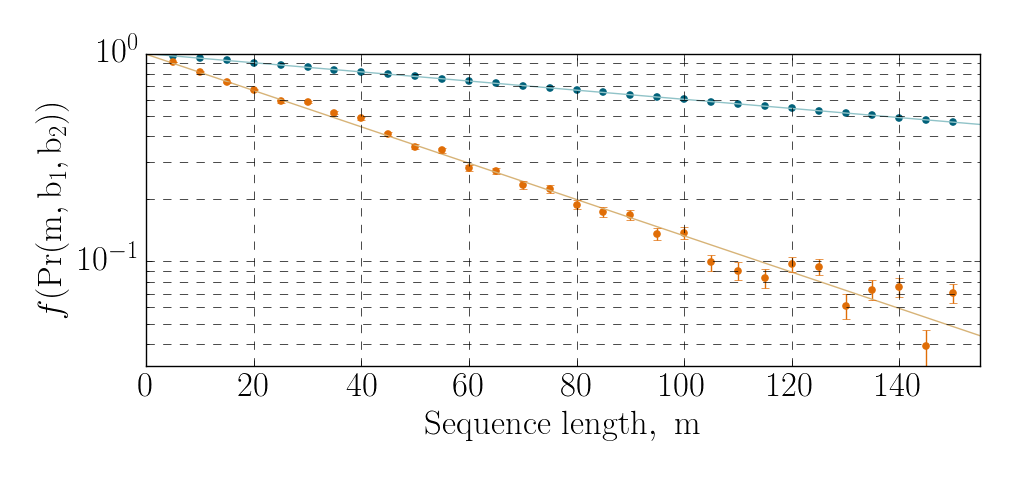}
		\caption{\label{fig:decay_curves}(Color online) Decay curves 
		corresponding to eqns.~\ref{eq:decay_curvep0} and 
		\ref{eq:decay_curvep1} for a standard 
			randomized benchmarking simulation with $A=\frac{1}{4}$ and $B=\frac{1}{2}$ respectively. 
			The shallow (blue) and steep (orange) lines correspond to 
			eqn.~\ref{eq:decay_curvep0} and eqn.~\ref{eq:decay_curvep1} 
			respectively.
			Each data point is obtained after averaging $500$ sequences of 
			fixed length. The noise model used to obtained
			this figure is described in the \emph{Numerical simulation} section. Fits to the data give an estimated value for average fidelity over $\mc D_8$ of $0.99257(9)$, which compares well with the true value of $0.9925$.}
	\end{figure}
	
	For our simulations, each operation within the dihedral group $\mc D_8$ is 
	generated by composing two gates; the first gate is in the subgroup 
	generated by $X$ and the phase gate $S=R_8(2)$, while the second 
	gate is either identity or the $T=R_8(1)$ gate. The error associated with 
	the first gate is a simple depolarizing channel with an average fidelity of 
	$0.9975$. For the second gate, the error arises only after the $T$ gate, 
	and corresponds to an over-rotation with an average fidelity of $0.99$. The 
	total average fidelity over $\mc D_8$ is $0.9925$. 
	Fig.~\ref{fig:decay_curves} shows the two decay
	curves described by eqns.~\ref{eq:decay_curvep0} and \ref{eq:decay_curvep1}. Appropriate fits
	lead to a precise estimate of $0.99257(9)$ for the average fidelity. 
	
\begin{figure}
		\centering
		\includegraphics[width= 1.00\linewidth]{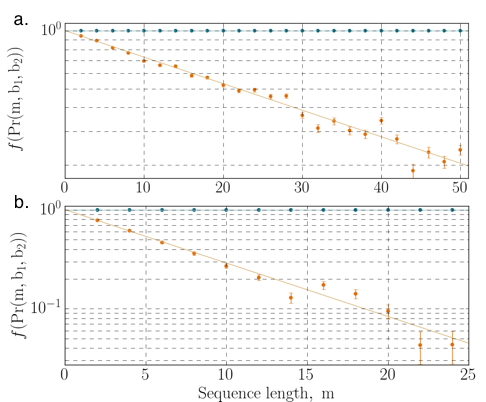}
		\caption{\label{fig:irb_curves} Decay curves corresponding to
			eqns.~\ref{eq:decay_curvep0} and \ref{eq:decay_curvep1} for an 
			interleaved	randomized benchmarking simulation with with 
			$A=\frac{1}{4}$ and $B=\frac{1}{2}$ respectively. The shallow 
			(blue) and steep (orange) lines correspond to 
			eqn.~\ref{eq:decay_curvep0} and eqn.~\ref{eq:decay_curvep1} 
			respectively. Each data point is obtained after averaging $500$ 
			sequences of fixed length. The noise model used to obtained
			this figure is described in the \emph{Numerical simulation} section.
			The top figure corresponds to high fidelity Clifford operations and a relatively noisy $T$ gate. The bottom figure corresponds to errors of the same magnitude on the Clifford operations and the $T$ gate. See text for details.}
	\end{figure}

	We also simulate the interleaved randomized benchmarking protocol in two different regimes 
	(see fig.~\ref{fig:irb_curves}). The first regime (fig.~\ref{fig:irb_curves}a) 
	corresponds to over-rotation errors that are small for the Clifford operations, with average fidelity $1- 10^{-6}$, but large for  the $T$ gate, 
	with average fidelity $1-10^{-2}$. The estimate of the fidelity of the $T$ 
	gate via our protocol is extremely precise in this regime: $0.9902(3)$. 
	The second regime (fig.~\ref{fig:irb_curves}b)  corresponds to a similar 
	over-rotation with average fidelity $0.99$ both for 
	the Clifford group and the $T$ gate. In this case the estimated value of $\mc{F}_{\rm{avg.}}(\mc{E}_T )$ is 
	$0.966$ and the implicit bound from
eqn. \ref{eq:implicit_bound} only guarantees $\mc{F}_{\rm{avg.}}(\mc{E}_T )$
	to lie the interval $[0.928, 0.995]$. The rather loose bound in this regime is an 
	open problem for interleaved randomized benchmarking and is not specific to the current protocol.

\emph{Conclusions.}---We have provided a protocol that extracts the average 
fidelity of the error arising over a group of single-qubit operations 
corresponding to the dihedral group. Of particular importance are $\mc{D}_8$ 
and $\mc{D}_4$, 
which enable efficient and precise benchmarking of the $T$ gate that plays a unique and important  role in leading 
proposals for fault-tolerant quantum computation. 

While we have explicitly assumed that the rotation axis is the $z$ axis, this 
is an arbitrary choice. Since any single-qubit unitary can be written as a 
rotation about some axis on the Bloch sphere, our protocol can be used to 
characterize any single-qubit gate.

The essence of this paper is to realize that the 2-design restriction
originally imposed in randomized benchmarking is too strict. Indeed, randomized 
benchmarking can be applied to any group whose Liouville representation 
contains few inequivalent irreps. Unraveling the class of all such algebraic 
structures is an open problem, though interesting groups such as $\langle CZ, 
X, R_j(1)\rangle$ abound, which would allow gates such as the $T$ gate to be 
characterized `in vivo' within an $n$-qubit circuit~\cite{Dugas15}.

\textit{Acknowledgments}--- This research was supported by the U.S. 
Army Research Office through grant W911NF-14-1-0103, CIFAR, the Government of 
Ontario, and the Government of Canada through NSERC and Industry Canada.

\bibliography{library}
\end{document}

%% file: macros.tex
\usepackage{hyperref,graphicx, amsmath, amssymb, amsthm, color, bm, bbm,dsfont}

\theoremstyle{plain}

\theoremstyle{definition}

\definecolor{nblue}{rgb}{0.2,0.2,0.7}
\definecolor{ngreen}{rgb}{0.1,0.5,0.1}
\definecolor{nred}{rgb}{0.8,0.2,0.2}
\definecolor{nblack}{rgb}{0,0,0}

\newcommand{\mc}[1]{\mathcal{#1}}

\newcommand{\bmc}[1]{\bm{\mathcal{#1}}}
\newcommand{\mbb}[1]{\mathbb{#1}}

\newcommand{\unit}{\mathbb{I}}

\newcommand{\ct}{^{\dagger}}
\newcommand{\av}[1]{\left\lvert #1\right\rvert}

\newcommand{\apa}[1]{\left(#1\right)}

\newcommand{\mat}[2]{\apa{\begin{array}{#1} #2\end{array}}}

\newcommand{\hidden}[1]{}